\documentclass[aps,prd,nofootinbib,tightenlines,notitlepage,floatfix,superscriptaddress,showkeys,onecolumn]{revtex4-1}

\RequirePackage{fix-cm}

\RequirePackage{color,graphicx,url}

\usepackage{placeins}

\let\Oldsection\section
\renewcommand{\section}{\FloatBarrier\Oldsection}

\let\Oldsubsection\subsection
\renewcommand{\subsection}{\FloatBarrier\Oldsubsection}

\let\Oldsubsubsection\subsubsection
\renewcommand{\subsubsection}{\FloatBarrier\Oldsubsubsection}

\usepackage{amsfonts,amsmath,amssymb,bm,xspace}
\usepackage{multirow}
\usepackage{bm}
\usepackage{enumitem}
\usepackage{nccmath}
\usepackage{xfrac}

\newcommand{\sat}{\mathrm{sat}}

\newcommand{\Lcal}{\mathcal{L}}
\newcommand{\psib}{\bar{\psi}}

\graphicspath{{./figures/}}

\usepackage[colorlinks=true]{hyperref}  

\usepackage{academicons} 
\newcommand{\orcid}[1]{\href{https://orcid.org/#1}{\textcolor[HTML]{A6CE39}{\aiOrcid}}}
\usepackage{xcolor}

\newcommand{\vac}{\mathrm{vac}}
\newcommand{\medium}{\mathrm{medium}}
\newcommand{\vacuum}{\mathrm{vacuum}}

\hypersetup{%
    pdfsubject=Paper,
    pdfkeywords={nuclear physics} {MBPT} {chiral EFT} {asymmetric nuclear matter}
    unicode=true,
    breaklinks=true,
     colorlinks   = true,
     linkcolor = blue,
     citecolor = blue,
     menucolor = blue,
     urlcolor = blue
}

\begin{document}

\title{Chiral confining Hartree-Fock Lagrangians based on Nambu-Jona--Lasino model}

\author{Mohamad Chamseddine}
\affiliation{Univ Lyon, Univ Claude Bernard Lyon 1, CNRS/IN2P3, IP2I Lyon, UMR 5822, F-69622, Villeurbanne, France \label{addr1}}

\author{J\'er\^ome Margueron}
\affiliation{Institut de Physique des 2 infinis de Lyon, CNRS/IN2P3, Universit\'e de Lyon, Universit\'e Claude Bernard Lyon 1, F-69622 Villeurbanne Cedex, France}
\affiliation{International Research Laboratory on Nuclear Physics and Astrophysics, Michigan State University and CNRS, East Lansing, MI 48824, USA} 

\author{Hubert Hansen}
\affiliation{Univ Lyon, Univ Claude Bernard Lyon 1, CNRS/IN2P3, IP2I Lyon, UMR 5822, F-69622, Villeurbanne, France \label{addr1}}

\author{Guy Chanfray}
\affiliation{Univ Lyon, Univ Claude Bernard Lyon 1, CNRS/IN2P3, IP2I Lyon, UMR 5822, F-69622, Villeurbanne, France \label{addr1}}

\begin{abstract}
We study a relativistic Hartree-Fock Lagrangian model which considers confinement, chiral symmetry breaking, nucleon form factor and short range correlations. The chiral potential originally based on the linear sigma-model is compared to an improved potential generated by the Nambu-Jona--Lasino (NJL) model for quark interaction. Our model is also anchored in fundamental hadronic properties predicted by Lattice-QCD calculations and a few nuclear empirical properties. We explore in a Bayesian approach the role of the saturation density, the energy per particle and the incompressibility modulus for the model selection. We find that most of our models could not reproduce these empirical quantities, unless a phenomenological "missing" energy is added. The properties of this "missing" energy are therefore inferred from our Bayesian analysis and we obtain that it shall be attractive. Finally we analyse the origin of the break down density in relativistic approaches and we relate it to the properties of the scalar potential. 
\end{abstract}

\maketitle

\section{Introduction}
\label{sec:intro}
The fundamental theory of strong interactions, Quantum Chromo-Dynamics (QCD), has shown great success in describing high-energy phenomena prevalent at extreme temperatures or densities~\cite{Weinberg}. In these regimes, the perturbative nature of QCD enables its application
from diagrammatic approaches.
However, when confronted with low energies, such as those encountered within nuclear physics -- where temperatures and densities are comparatively low (below $1$ GeV.fm$^{-3}$) -- QCD transitions into a non-perturbative regime. This non-perturbative nature manifests itself through non-linear
phenomena like the spontaneous breaking of symmetries endowing the QCD vacuum with a non-trivial structure, and a seemingly unrelated property, the color confinement.

In the low-energy regime, where direct perturbative approaches fall short, several approaches capturing the symmetries inherent in the theory have been proposed. Notably, the Nambu-Jona--Lasinio (NJL) model~\cite{Nambu:1961fr} and the Chiral Effective Field Theory ($\chi$-EFT)~\cite{Weinberg} have emerged as promising frameworks.

In a previous paper~\cite{Chamseddine}, we have developed a mean-field approach for nuclear matter based on a chiral version -- as originally proposed in \cite{Chanfray2005,Chanfray2007,Massot2008,Massot2009,Chanfray2011} -- of the relativistic theories of Walecka and collaborators \cite{SerotWalecka1986,Walecka1997} in which we have included chiral symmetry breaking, confinement, nucleon finite size through form factors (FF) and short range correlations (SRC) due to the repulsive nature of the nuclear interaction at short distance. The model employed, the so-called relativistic Hartree-Fock with chiral symmetry and confinement (RHF-CC), constitutes a specific implementation of a theoretical framework called the chiral confining model, see a concise summary in Ref.~\cite{Chanfray-Schuck}. It  offered advantageous features which are further detailed hereafter:
\begin{enumerate}
\item A modern understanding of the
"nuclear physics" scalar sigma meson $\sigma_W$ employed in Walecka model, which has been a subject of controversy. This scalar field is identified with the chiral invariant field $s=S-F_\pi$, associated to the radial fluctuation of the chiral condensate $S$ around the ”chiral radius” $F_\pi$, which is the pion decay constant \cite{Chanfray2001}.
\item An enhanced saturation mechanism based on the so-called Walecka mechanism (the competition between scalar and vector fields) and completed with the interplay between the tadpole term of the chiral potential and the polarisation term originating from quark confinement \cite{Chanfray2005,Chanfray2007,Massot2008,Massot2009}.
\end{enumerate}

To clarify the first point, let us consider the lightest $u$ and $d$ quarks with the flavor number $N_f =2$. The chiral field associated to the fluctuations of the quark condensate $\langle \bar q q\rangle$ resulting from chiral symmetry breaking can be parameterised in terms of a $\rm{SU}(2)$ matrix $M$ as:
\begin{eqnarray}
M&=&\sigma + i\vec{\tau}\cdot\vec{\pi}\equiv S\, U\equiv (s\, +\, F_{\pi})\,U\equiv (\sigma_W +\, F_{\pi})\,U \, ,
\label{REPRES}
\end{eqnarray}
with $U(x)=e^{i\,{\vec{\tau}\cdot\vec{\phi}(x)}/{F_\pi}}$. The scalar field $\sigma$ (respect. $S$) and pseudo-scalar fields $\vec{\pi}$ (respect. $\vec{\phi}$) written in Cartesian (respect. polar) coordinates appear as in-medium dynamical degrees of freedom and may deviate from the vacuum value, $\left\langle \sigma\right\rangle_{\vac}=\left\langle S\right\rangle_{\vac} = F_\pi\propto\left\langle \overline{q}q\right\rangle_{\vac}$. The sigma and the pion in polar coordinates, associated to the amplitude fluctuation $s\equiv\sigma_W$ and the phase $\vec{\phi}$ of this condensate, are considered in our approach as effective degrees of freedom. Their dynamics are governed by an effective chiral potential, $V_\chi\left(s,\vec{\phi}\right)$, having a typical Mexican hat shape associated with the broken chiral symmetry of the QCD vacuum. Again, one may be tempted to identify $\sigma_W$ with the scalar field $\sigma$ in the Cartesian coordinates. In this case, terms of order $m_\pi$ will appear in the NN interaction, which is not allowed as first argued by Birse~\cite{Birse94}. 

One notable challenge arises when using chiral effective theories (without confinement) to describe nuclear saturation \cite{Boguta83,KM74,BT01,C03}. Regardless of the specific chiral model employed, these systems are unstable and collapse. This is due to the in-medium reduction of the sigma mass: the in-medium value of $S$, $S_{\medium}$, differs from its value in vacuum, $S_{\vacuum}$, which corresponds to the minimum of the "Mexican hat" vacuum effective potential. 
The sigma mass being proportional to $V_\chi''(S)$ and $V_\chi''(S_{\medium}) < V_\chi''(S_{\vacuum})$, leads to its reduction. This effect can be attributed to the presence of a $s^3$ tadpole diagram in the potential $V_\chi$, generating attractive three-body forces that disrupt saturation even when the repulsive three-body force from the Walecka mechanism is present. This brings us to the second point.

So far in the discussion, we haven't employed the knowledge from QCD that nucleons are composed of quarks and gluons. 
Consequently, nucleons respond to the presence of the surrounding nuclear scalar field by readjusting their quark sub-structure distribution,
as shown in the pioneering work of P. Guichon~\cite{Guichon1988} in the quark-meson coupling  (QMC) model. As proposed in \cite{Chanfray2005,Chanfray2007,Massot2008,Massot2009}, this response can be accounted for by introducing the nucleon's reaction to the scalar field $s$ through the nucleon scalar susceptibility $\kappa_\mathrm{NS}=d^2M_N^\star(s)/ds^2$ where the nucleon mass $M_N^\star(s)$ represents the nucleon polarised mass as defined in Eq.~\eqref{eq:nucleon_mass}. In our previous papers we have introduced a dimensionless parameter we called $C$, that we call $C_\mathrm{NS}$ for convenience in this work:
\begin{equation}
C_\mathrm{NS} = \frac{\kappa_\mathrm{NS} F^2_\pi}{2M_N} \, .  \label{eq:coefC}
\end{equation}
It happens that the quadratic dependence of the mass $M_N^\star(s)$ on the scalar field $s$ generates a repulsion which counterbalances the role of the attractive tadpole term responsible for the destruction of saturation.  

It seems that the issue of the collapse induced by the chiral potential is solved by the contribution of the nucleon polarisation, itself due to the quark sub-structure of nucleons. However, there is still an issue in this solution which we now discuss.
As pointed out in our recent work \cite{Chanfray_Cchi}, this new issue is about the numerical values taken by the dimensionless parameter $C_\mathrm{NS}$.
The MIT bag model in the QMC framework predicts $C_\mathrm{NS} \sim 0.5$, or the QCD-connected version of the chiral confining model in Ref.~\cite{chanfray-universe,Chanfray-Schuck} predict $C_\mathrm{NS} \sim 0.3$.
However Lattice-QCD (LQCD) data from the Adelaide group~\cite{LWY2003,Leinweber2004,TGLY04,Armour_2010} favors values larger than one, and in our previous work employing a Bayesian analysis, we predict a value for $C_\mathrm{NS}$
around $C_\mathrm{NS} \approx 1.4$~\cite{Chamseddine}.
So far, our studies have employed the Linear Sigma Model (L$\sigma$M) for the chiral effective potential~\cite{Chanfray2007,Massot2008,Massot2009,Massot2012,Rahul2022,Chamseddine}, given by:
\begin{equation}
V_{\chi,\mathrm{L\sigma M}}(s)=\frac{1}{2}\,m^2_s s^2\, +\,\frac{1}{2}\frac{m^2_s -m^2_\pi}{ F_\pi}\, s^3\,+\,
\frac{1}{8}\,\frac{m^2_s -m^2_\pi}{ F^2_\pi} \,s^4 \label{eq:VLSM}\, ,
\end{equation}
where we see the cubic tadpole term mentioned previously, which is responsible for the collapse. The usage of such a model is probably too naive, and a richer NJL potential has recently been suggested in Ref.~\cite{Chanfray_Cchi} based on a more accurate description of the low-energy realisation of chiral symmetry in the hadronic sector.

In this paper, we investigate the effect of this NJL potential on the properties of symmetric nuclear matter in a similar manner to what was done in Ref.~\cite{Chamseddine} for the RHF-CC version of the chiral confining model.
We first introduce this new NJL potential in Sec.~\ref{sec: NJL potential intro} and compare it to the L$\sigma$M one. In Sec.~\ref{sec:RHF-CC} we inject this new potential into our RHF-CC calculation, and we perform a Bayesian analysis, which is introduced in Sec.~\ref{sec:bayesian}, to see what values of the NJL parameters are able to reproduce the desired nuclear empirical parameters (NEP). Throughout this work, we address two questions: one relating to the lack of attraction under our considered priors, inducing a "missing energy" term, as discussed in Sec.~\ref{sec:missing energy}, and the other one relating to a the breakdown of the model at higher densities in Sec.~\ref{sec:breakdown}.

\section{The NJL chiral confining model}
\label{sec: NJL potential intro}

The main idea to construct the NJL chiral potential is to consider nuclear matter as made of nucleons, seen as Y-shaped strings with quarks at the ends
subject to a non-perturbative confining force~\cite{Chanfray2011}. The constituent NJL quarks acquire significant mass from the quark condensate, which serves as the order parameter associated with the spontaneous breaking of chiral symmetry in the QCD vacuum. As the density $n$ of nuclear matter increases, the presence of nucleons alters the QCD vacuum, leading to a decrease in the quark condensate and a gradual restoration of chiral symmetry. The explicit construction of the chiral-invariant scalar field $\mathcal{S}$, associated with the radial fluctuation mode of the chiral condensate, in the NJL framework has been done in Ref.~\cite{Chanfray2011}.  It follows that the mass of the constituent quarks aligns with the in-medium expectation value, $M=\overline{\mathcal{S}}(n)$. To characterize this field in the context of nuclear physics, we define an "effective" or "nuclear physics" scalar field $s$ by scaling the chiral-invariant scalar field $\mathcal{S}$ as follows:
\begin{equation}
\mathcal{S} \equiv\frac{M_0}{F_\pi}\,S\equiv\frac{M_0}{F_\pi}\,\left(s+F_\pi\right) \label{eq:Zfactor}
\end{equation}
Here, $M_0=\overline{\mathcal{S}}(s=0)$ represents the constituent quark mass in vacuum. The vacuum expectation value of the "effective" scalar field, $\overline{S}=F_\pi$, coincides by construction with the value of the pion decay constant ${F_\pi}$ and we remind that its fluctuating component, $s$, is suggested to be identified to $\sigma_W$. For a detailed exposition of this construction, we refer to Refs.~\cite{Chanfray2011, Chanfray_Cchi,chanfray-universe}. Note that from Eq.\eqref{eq:Zfactor}, one can deduce,
\begin{equation}
\frac{\partial}{\partial s}  = \frac{M_0}{F_\pi}\,\frac{\partial}{\partial \mathcal{S}}
\end{equation}

\subsection{The NJL chiral potential}

We define the following NJL Lagrangian:
\begin{eqnarray}
{\cal L}_\mathrm{NJL}&=& \overline{\psi}\left(i\,\gamma^{\mu}\partial_\mu\,-\,m\right)\,\psi\,+\,\frac{G_1}{2}\,\left[\left(\overline{\psi}\psi\right)^2\,+\
\left(\overline{\psi}\,i\gamma_5\vec\tau\,\psi\right)^2\right]\nonumber\\
& &-\,\frac{G_2}{2}\,\left[\left(\overline{\psi}\,\gamma^\mu\vec\tau\,\psi\right)^2\,+\,
\left(\overline{\psi}\,\gamma^\mu\gamma_5\vec\tau\,\psi\right)^2\,+\,\left(\overline{\psi}\,\gamma^\mu\,\psi\right)^2\right]. \label{LNJL}
\end{eqnarray}
where $\psi$ are the quark fields, and there are four parameters: the coupling constants $G_1$ (scalar) and $G_2$ (vector), the current quark mass $m$ and a (non-covariant) cutoff parameter $\Lambda$. We note that in the QCD-connected version of the chiral confining model, the $G_1$ parameter behaves as $G_1\sim\sigma T^4_g$ where $\sigma$ is the string tension and $T_g$ is the string width, whereas the cutoff $\Lambda$ is related to the inverse of the string width, see Ref.~\cite{chanfray-universe,Chanfray-Schuck} for further details. 

The parameter $G_2$ however cannot be fixed by a physical observable. In the following, we therefore consider two characteristic choices for $G_2$ coupling: i) $G_2 =0$ corresponding to the absence of vector interaction, and ii) $G_2=G_1$ compatible with the $\pi$-$a_1$ mixing, see Refs.~\cite{Chanfray2011, Chanfray_Cchi,chanfray-universe}. The remaining three parameters ($G_1$, $m$ and $\Lambda$) are determined to reproduce the pion mass $M_\pi$ and the pion decay constant $F_\pi$. Since there is one more parameter than the number of constraints, we expect to obtain a family of solutions, see Section~\ref{sec:Parametrisation} for more details.

The NJL chiral potential can be obtained from the NJL Lagrangian after using path integral techniques and semi-bosonisation, see details in Ref.~\cite{Chanfray2011}, and can be 
expressed as:
\begin{equation}
V_{\chi,\mathrm{NJL}}(S)=-2N_c N_f\,\big(I_0(\mathcal{S})\,-\,I_0(M_0)\big) \,+\,\frac{\left(\mathcal{S}-m\right)^2 -\left(M_0 - m\right)^2}{2\,G_1}.\label{eq:full_VNJL}
\end{equation}
Here, the term $-2N_c N_f I_0(\mathcal{S})$ corresponds to the total in-medium energy of the Dirac sea of constituent quarks, with $I_0(\mathcal{S})$ denoting the NJL loop integral given below. The vacuum constituent quark mass $M_0$ corresponds to the minimum of the chiral effective potential, satisfying $V'_{\chi,\mathrm{NJL}}(s=0)=0$, where $V'$ represents the derivative with respect to the scalar field $s$. $M_0$ can be obtained by solving the gap equation:
\begin{equation}
M_0 = m\,+\,4N_c N_f M_0\,G_1\,I_1(M_0),    \label{eq:GAP}
\end{equation}
In Eq.~\eqref{eq:GAP}, $I_1(M_0)$ denotes another NJL loop integral. The various loop integrals encountered here are given by:
\begin{eqnarray}
&& I_0(\mathcal{S})=\int_0^\Lambda \frac{d{\bf p}}{(2\pi)^3}\,E_p(\mathcal{S}),\quad	I_1(\mathcal{S})=\int_0^\Lambda \frac{d{\bf p}}{(2\pi)^3}\,\frac{1}{2\,E_p(\mathcal{S})},\nonumber\\
&& I_2(\mathcal{S})=\int_0^\Lambda \frac{d{\bf p}}{(2\pi)^3}\,\frac{1}{4\,E^3_p(\mathcal{S})},	\quad J_3(\mathcal{S})=\int_0^\Lambda\frac{d{\bf p}}{(2\pi)^3}\,\frac{3}{8\,E^5_p(\mathcal{S})} , \nonumber\\
&& J_4(\mathcal{S}) = \int_0^\Lambda\frac{d{\bf p}}{(2\pi)^3}\,\frac{15}{16\,E^7_p(\mathcal{S})} \, , \label{PARAM1}
\end{eqnarray}
where $E_p(\mathcal{S})=\sqrt{\mathcal{S}^2 + p^2}$. These integrals depend on the scalar field $\mathcal{S}$ and are used in the evaluation of the gap equation.

\subsection{Expansion of the NJL chiral potential in powers of the $s$ field}
\label{sec:Linearisation}
To compare between the NJL chiral potential and the L$\sigma$M, we perform an expansion of the NJL potential in Eq.~\eqref{eq:full_VNJL} around the vacuum ($s = 0$) and we get:
\begin{equation}
V_{\chi,\mathrm{NJL}}(s)\approx  \frac{1}{2}m^2_s {s}^2 +\frac{1}{2}\frac{m^2_s -m^2_\pi}{ F_\pi} \big(1-C_{\chi,\mathrm{NJL}}\big){s}^3 
+\frac{1}{8}\frac{m^2_s -m^2_\pi}{ F_\pi^2} \big(1-6~C_{\chi,\mathrm{NJL}} + D_{\chi,\mathrm{NJL}}\big) {s}^4
+...,
\label{eq:vchiNJL}
\end{equation}
where the canonical pion mass $m_\pi=\sqrt{\frac{m M_0}{G_1 F^2_\pi}}$ is calculated in the bosonized NJL model. The effective scalar mass $m_s$ is defined as
\begin{equation}
m^2_s=4\,M^2_0 \, \frac{f^2_\pi}{F^2_\pi} + \,m^2_\pi\, , 
\label{eq:msigma}
\end{equation}
with
\begin{equation}
\label{eq:fpi}
f^2_\pi = \frac{F^2_\pi}{1-4G_2 F^2_\pi} = 2N_c N_f M_0^2 I_2(M_0) \, .
\end{equation}
We refer the reader to Ref.~\cite{Chanfray2011} for more details on the derivation of the previous equations.
The parameters $C_{\chi,\mathrm{NJL}}$ and $D_{\chi,\mathrm{NJL}}$ are defined as:
\begin{eqnarray}
C_{\chi,\mathrm{NJL}}&=&\frac{2}{3}\,\frac{M^2_0\,J_3(M_0)}{I_2(M_0)}, \\ 
D_{\chi,\mathrm{NJL}}&=&\frac{4}{3}\,\frac{M^4_0\,J_4(M_0)}{I_2(M_0)}\, . 
\end{eqnarray}

The parameters $C_{\chi,\mathrm{NJL}}$ and $D_{\chi,\mathrm{NJL}}$ represent the correction to the original L$\sigma$M potential~\eqref{eq:VLSM} induced by the NJL model at orders three (tadpole term) and four. The parameter $C_{\chi,NJL}$ for instance modifies the cubic term which softens its attractiveness.
At the limit $C_{\chi,NJL} = 1$, the cubic tadpole term becomes zero, as in the case of the QMC model \cite{Guichon1988,Guichon2004}, while at the limit $C_{\chi,NJL} = 0$ the NJL potential is identical to the L$\sigma$M potential (at order three).

\begin{figure}[t]
\centering
\includegraphics[width=0.5\textwidth]{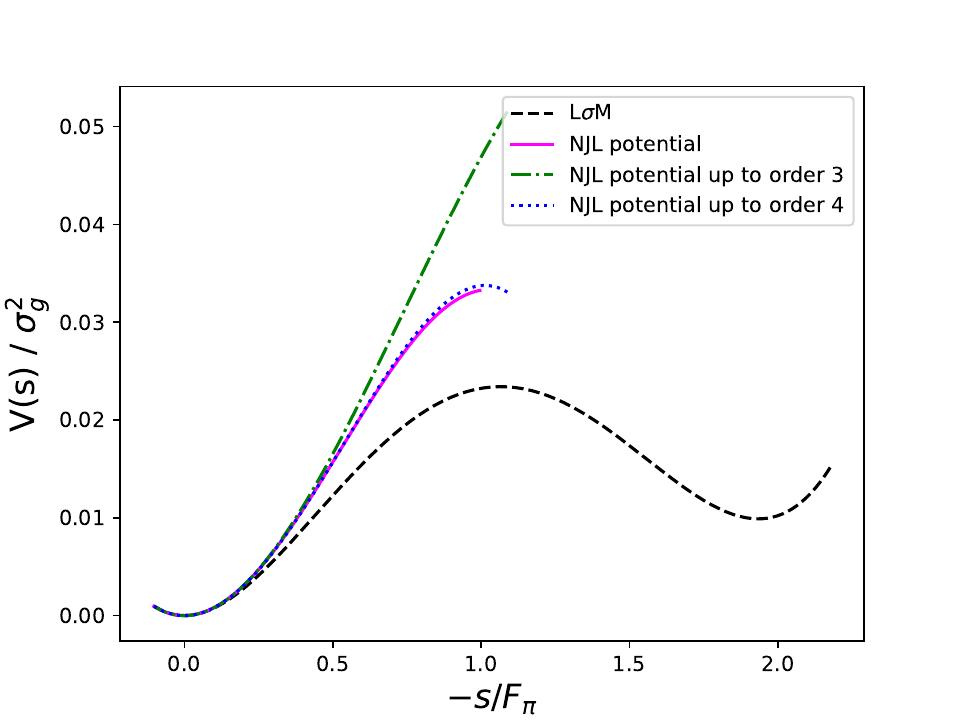}
\caption{The L$\sigma$M and NJL potentials (in units of the string tension $\sigma_g^2$, where $\sigma_g=0.18$ GeV$^2$) are shown as a function of the scalar field $-s / F_\pi$.
The L$\sigma$M (dashed line) is given by Eq.~\eqref{eq:VLSM}, the NJL model (full line) by Eq.~\eqref{eq:full_VNJL}, 
and the NJL potential up to order three (dash-dotted line) and four (dotted line) and given by Eq.~\eqref{eq:vchiNJL}. The potentials are plotted for the following parameters: $m = 5.9$~MeV, $\Lambda = 586$~MeV, $G_1 = 14.21$~GeV$^{-2}$, $m_\pi = 139.6$~MeV, $F_\pi = 91.9$~MeV, $M_0 = 398$~MeV, and $m_s = 809$~MeV. Note graphically the role of the explicit symmetry breaking term, $-F_\pi m_\pi^2 s$, which breaks the degeneracy between the two minimums of the "Mexican hat" potential.}
\label{fig:potential_comparison}
\end{figure}

A comparison of the L$\sigma$M and NJL potentials is shown in Fig.~\ref{fig:potential_comparison} for the parameters given in the caption, $m = 5.9 $~MeV, $\Lambda = 586$~MeV, $G_1 = 14.21$~GeV$^{-2}$, which fix $m_\pi$ and $F_\pi$ to their centroid values given in Tab.~\ref{tab:fitting}. From Eqs.~\eqref{eq:GAP} and \eqref{eq:msigma}, we obtain $M_0 = 398$~MeV and $m_s = 809$~MeV. The two minimums of the L$\sigma$M potentials are not degenerate because of the explicit symmetry breaking due to the pion mass and incorporated in Eq.~\eqref{eq:VLSM}, more details are given in the introduction of Ref.~\cite{Chamseddine} for instance.

Fig.~\ref{fig:potential_comparison} shows that the NJL potential and its approximation up to fourth order in $s$ agrees very well up to the chiral restoration at $s = -F_\pi$, which represents the upper limit for the scalar field considered in our model. The L$\sigma$M and the 
NJL potential coincides for low values of the scalar field $s$ and differences are visible for $s\gtrsim -0.3F_\pi$. The cubic approximation reproduces well the NJL potential up to about $s\approx -0.6F_\pi$, while the quadratic approximation is almost on top of NJL potential up to chiral restoration $s= -F_\pi$.

In the following work we will use the NJL potential~\eqref{eq:full_VNJL}, and we set for simplicity in the notations $C_\chi\equiv C_{\chi,\mathrm{NJL}}$, $D_\chi\equiv D_{\chi,\mathrm{NJL}}$ and $V_\chi\equiv V_{\chi,\mathrm{NJL}}$.

\section{Relativistic Hartree-Fock approach including NJL chiral potential and confinement}
\label{sec:RHF-CC}

The Lagrangian describing the relativistic Hartree-Fock approach with chiral potential and confinement (RHF-CC) can be written as the sum of a kinetic fermionic term,
\begin{equation}
\label{eq:L_kinetic}
\Lcal_\psi = \psib \left( i \gamma^{\mu} -M_N \right)\partial_{\mu} \psi \, ,
\end{equation}
where the field $\psi$ represents the nucleon spinor, and of the meson-nucleon term,
\begin{equation}
\Lcal_{m} = \Lcal_{s} + \Lcal_{\omega} + \Lcal_{\rho} + \Lcal_{\delta} + \Lcal_{\pi} \, ,
\label{eq:L_mesons}
\end{equation}
collecting all meson contributions considered in our model: $s$, $\omega$, $\rho$, $\delta$ and $\pi$. Note that the term associated with the scalar field $s$ contains also the dynamical chiral potential $V_\chi$. Using notation of Ref.~\cite{Massot2008} these meson-nucleon terms can be expressed as,
\begin{align}
\label{eq:L_meson}
\Lcal_s =& \big(M_N - M^\star_N(s)\big) \bar{\psi} \psi - V_\chi(s) + \frac{1}{2} \partial^\mu s \partial_\mu s\, , \nonumber \\
\Lcal_\omega =& -g_\omega \omega_\mu \bar{\psi} \gamma^\mu \psi + \frac{1}{2} m_\omega^2 \omega^\mu \omega_\mu - \frac{1}{4} F^{\mu \nu} F_{\mu \nu}\, , \nonumber \\
\Lcal_\rho =& -g_\rho \rho_{a \mu} \bar{\psi} \gamma^\mu \tau_a \psi + g_\rho \frac{\kappa_\rho}{2 M_N} \partial_\nu \rho_{a \mu} \bar{\psi} \sigma^{\mu \nu} \tau_a \psi \nonumber \\
&+ \frac{1}{2} m_\rho^2 \rho_{a \mu} \rho_a^\mu - \frac{1}{4} G_a^{\mu \nu} G_{a \mu \nu}\, , \\
\Lcal_\delta =& -g_\delta \delta_a \bar{\psi} \tau_a \psi - \frac{1}{2} m_\delta \delta_a \delta_a + \frac{1}{2} \partial^\mu \delta_a \partial_\mu \delta_a\, , \nonumber \\
\Lcal_\pi =& \frac{g_A}{2 F_\pi} \partial_\mu \varphi_{\pi a} \bar{\psi} \gamma^\mu \gamma^5 \tau_a \psi - \frac{1}{2} m_\pi^2 \varphi_{\pi a} \varphi_{\pi a} \nonumber \\
&+ \frac{1}{2} \partial^\mu \varphi_{\pi a} \partial_\mu \varphi_{\pi a}\, , \nonumber
\end{align}
where the symbols have their usual meaning. Historically, the pion coupling was defined as $\tilde{F}_\pi/m_\pi$, which is replaced by $g_A/(2F_\pi)$ considering $\tilde{F}_\pi=m_\pi g_A/(2 F_\pi)$, where $g_A$ is the axial coupling constant and $F_\pi$ is the pion decay constant.
This Lagrangian only differ from the one in Ref.~\cite{Chamseddine} by the use of the NJL chiral potential $V_\chi(s)$ instead of $V_{L \sigma M}(s)$. The leading order effect induced by the quark confinement is incorporated in the nucleon mass through the nucleon polarisability $\kappa_\mathrm{NS}$ as in the QMC model: 
\begin{equation}
M^\star_N(s)=M_N+g_s s+\frac{1}{2}\kappa_\mathrm{NS}\left(s^2+\frac{s^3}{3F_\pi}\right).
\label{eq:nucleon_mass}
\end{equation}
The term in $s^3$ is introduced as in Ref.~\cite{Chanfray2005} such that the susceptibility vanishes at chiral restoration. It is the simplest way to incorporate chiral restoration, however it is not impossible that this prescription is too simple and the real evolution of the mass close to chiral restoration may be more complex. The present model is indeed expected to be valid only for small polarisation of the nucleon and may break-down well before the chiral restoration. The limit at chiral restoration shall however be satisfied in all cases. One can estimate the domain where the nucleon mass is well described by Eq.~\eqref{eq:nucleon_mass}. Fixing this limit 
at
the values where the highest order correction is larger than 10-20\% of the previous order, one finds $s_{\max} \approx -0.3F_\pi$ to $-0.6F_\pi$, which also corresponds to the break-down limit of our model, see Sec.~\ref{sec:breakdown}. We can conclude that our model can safely be employed up to the break down density.

The in-medium susceptibility is defined as:
\begin{equation}
\label{eq:kappa_tilde}
\tilde\kappa_\mathrm{NS}(s)={\partial^2M^\star_N(s)\over\partial s^2}=
\kappa_\mathrm{NS}\left(1+{s\over f_\pi}\right) \, ,
\end{equation} 
which indeed vanishes at full chiral restoration, i.e., $\bar s=-F_\pi$, where $\bar s$ is the mean value taken by the field $s$ in the ground state.

The nuclear matter energy density is defined as~\cite{Chamseddine} 
\begin{equation}
\varepsilon_0= \epsilon^K + \epsilon^H_{s+\omega+\rho+\delta} + \epsilon^F_{s+\omega+\rho+\delta+\pi}
\end{equation}
where $\epsilon^K, \epsilon^H$ and $\epsilon^F$ refer to the kinetic, Hartree and Fock contributions respectively.
The Fock term includes all finite-size and short-range correlations generated by the monopole form factors (FF) and the Jastrow ansatz parameterised by $\mathrm{q_c}$, which corresponds to model D of Ref.~\cite{Chamseddine}.

The Adelaide group~\cite{Leinweber2004,LWY2003,TGLY04,Armour_2010} have shown that the nucleon mass can be expanded in terms of the pion mass squared, $m^2_\pi$, in the following form:
\begin{equation}
\label{eq:LQCD_MN}
    M_N(m^2_\pi) = a_0 + a_2 m^2_\pi + a_4 m^4_\pi + ... + \Sigma_\pi(m^2_\pi, \Lambda)\, ,
\end{equation}
where the pionic self-energy $\Sigma_\pi(m^2_\pi, \Lambda)$ containing non analytical contribution is explicitly separated out and they have determined the values of the parameters $a_i$ from LQCD.
Using Eq.~\eqref{eq:LQCD_MN}, we can relate $a_2$ and $a_4$ to our parameters via the following relations (see Ref.~\cite{Chanfray_Cchi}):
\begin{eqnarray}
a_2 &=& \frac{F_\pi\, g_{S}}{m^2_s} \, 
\label{eq:a2} \\
a_4 &=&-\frac{F_\pi\,g_{S}}{2 m^4_s}\,\left(3\,-\,2\,\tilde{C}_L \right)
\label{eq:a4}
\end{eqnarray}
with $\tilde{C}_L = M_N/(g_S\,F_\pi)\,C_\mathrm{NS}\,+\,3/2 \,C_\chi$.
Again, note that when $C_\chi = 0$, we find the same LQCD relations as for the L$\sigma$M in Refs.~\cite{Massot2008,Chanfray2007,Chamseddine}.

In our study of the RHF-CC model, we have used a Bayesian approach to explore the model parameters phase space, and to translate uncertainties from our constraints to model parameter uncertainties. In the next section we briefly describe the Bayesian approach and the various nomenclatures encountered.

\subsection{Bayesian analysis}
\label{sec:bayesian}

\begin{table}[t]
\tabcolsep=0.33cm
\def\arraystretch{1.5}
\caption{\label{tab:fitting}%
The values of the
constraining quantities:
the parameters $a_2$ and $a_4$ from LQCD, the parameters $F_\pi$, $M_\pi$, $M_0$, $\Lambda^2 G_1$ from the NJL sector, and the NEP $E_{\sat}$, $n_{\sat}$ and $K_{\sat}$. In the last column, the symbol $\bigstar$ refers to the quantities for which we impose a uniform distribution (since we are agnostic in the distribution of these quantities), while for the rest a Gaussian distribution is considered. Note that in our preliminary study we do not consider the constraint on $M_0$.}
\begin{tabular}{cccc}
\hline
Parameters & Ref. & centroid & std. dev. (or total width)\\
\hline
$a_2$ (GeV$^{-1}$) &  \cite{LWY2003} & 1.553 & 0.136$\bigstar$  \\
$a_4$ (GeV$^{-3}$) &  \cite{LWY2003} & -0.509 & 0.054$\bigstar$ \\
$F_\pi$ (MeV) & & 92 & 2 \\
$m_\pi$ (MeV) & & 139.6 & 2 \\
$M_0$ (MeV) & & $< 400$ MeV \\
$\Lambda^2 G_1$ & & 5.5 & 9$\bigstar$ \\
$E_{\sat}$ (MeV) & \cite{Margueron2018} & -15.8 & 0.3\\
$n_{\sat}$ (fm$^{-3}$) & \cite{Margueron2018} & 0.155 & 0.005\\ 
$K_{\sat}$ (MeV) & \cite{Margueron2018} & 230 & 20\\
\hline
\label{tab:modelparameterfit}
\end{tabular}
\end{table}
 
In the Bayesian approach, the so-called "data" are the quantities which are constrained by
experimental data, e.g., $F_\pi$, $m_\pi$, and the NEPs $n_\sat$, $E_\sat$ and $K_\sat$, as well as by the analysis of well established theories, e.g., the LQCD parameters $a_2$ and $a_4$. The "model" parameters are quantities like coupling constants, e.g., $g_\omega$, $g_s$ and $C_\mathrm{NS}$, or masses, e.g. $m_s$, which directly enter into the definition of the model. A "model" can therefore can be simply represented by the set $ \{ \theta_i \} $ of these "model" parameters. They will be fixed in such a way as to reproduce the previously mentioned "data". 
The Bayesian probability associated to a set of models and to a set of data can be obtained in two different ways since (Bayes theorem):
\begin{equation}
P(\{\theta_i\} \mid \textrm{data}) \times P(\textrm{data}) =  P(\textrm{data}\mid \{\theta_i\})\times P(\{\theta_i\}),
\label{eq:bayes}
\end{equation}
where $P(\{\theta_i\} \mid \textrm{data})$ is the posterior probability associated to a model given a set of "data", $P(\textrm{data})$ is the evidence which is defined as the normalisation of the posterior probability.
We have $P(\textrm{data}\mid \{\theta_i\})$ the likelihood probability associated to a set of "data" given a model, which can be expressed for instance as
\begin{equation}
\log P(\textrm{data}\mid \{\theta_i\}) = -\frac 1 2 \frac{1}{N_\mathrm{dof}}\sum_i  \frac{\left[O_i(\mathrm{data})-O_i(\{\theta_i\})\right]^2}{\Delta O_i^2}  \, ,
\label{eq:likelihood}
\end{equation}
where $N_\mathrm{dof}$ is the number of independent "data" -- the number of "data" minus the number of "model" parameters -- and the sum goes over the number of data where $O_i(\mathrm{data})$ represents a data and $O_i(\{\theta_i\})$ its prediction from the model and $\Delta O_i$ is the uncertainty in the data originating from the uncertainty in the measurement, but it can also incorporate a systematic uncertainty estimation associated to the "model" (in this case one shall consider the sum of the square of the different uncertainties~\cite{Dobaczewski:2014}). Finally, the probability $P(\{\theta_i\})$ in Eq.~\eqref{eq:bayes} is the prior probability, which represents the \textsl{a-priori} knowledge on the model parameters. For priors, Gaussian distributions are considered when dealing with experimental data since the Gaussian width can be associated to the experimental standard deviation (std. dev.), and flat distributions are taken when a width is not meaningful, e.g., for the LQCD data which are extracted from \cite{LWY2003}. The constraining quantities which are considered in our study are given in Tab.~\ref{tab:fitting}.

\subsection{Parameters for the NJL model}
\label{sec:Parametrisation}

In the following preliminary study, we explore the values taken by the NJL model parameters, which are $G_1$, $m$ (or equivalently $M_0$) and $\Lambda$, conditioned only by 
the "data": $m_\pi$ and $F_\pi$ as given in Tab.~\ref{tab:fitting}. We also add an extra prior requiring a "natural" value of $\Lambda^2 G_1$ in the following sense: a large value of $\Lambda^2 G_1$ would imply an unrealistic dominance of the interaction over the kinetic contributions whereas a small value would mean a perturbative interaction, hence forbidding spontaneous chiral symmetry breaking. We would thus like to have $\Lambda^2 G_1$ in the range $ \sim 1-10$, and we consider a flat distribution since no values in this range has any reason to be favored over the others. 
With these in mind, we can run a Bayesian analysis that fully explores the phase space of the parameters. In the following, we only show the case $G_2 = 0$, since the other case $G_2 = G_1$ is qualitatively similar.

\begin{figure}
\centering
\includegraphics[width=0.9\textwidth]{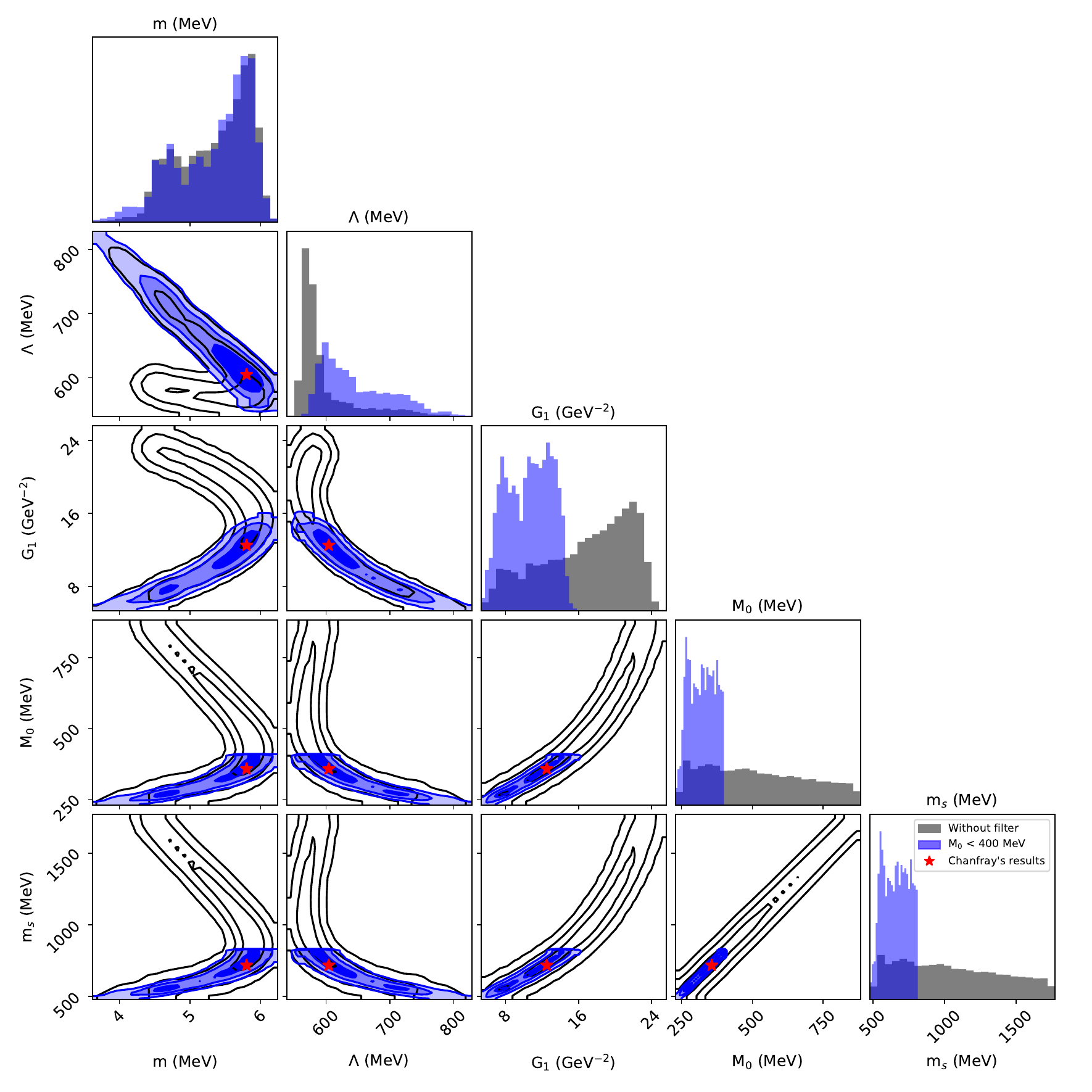}
\caption{Corner plot showing the posterior probabilities and the correlations between the various NJL parameters for $G_2=0$ (grey band). The valid range of solutions extends to high values of $M_0$ which are considered unphysical per Ref.~\cite{Biguet2015}. The blue color indicates the distributions after we cut out these unphysical values (i.e imposing $M_0 < 400$ MeV). The star marks the model parameters considered in Ref.~\cite{Chanfray_Cchi,chanfray-universe,Chanfray-Schuck}.}
\label{fig:NJL_only}
\end{figure}

We represent in Fig.~\ref{fig:NJL_only} the posterior probability obtained from the Bayesian analysis in the form of a corner plot, where we place on the diagonal the posterior probabilities marginalised over all parameters but one, and the correlations between the parameters are given in the off-diagonal insets. The result of the preliminary study is shown in gray. The most noticeable feature is that a wide range of valid solutions are obtained, instead of single peaked ones. This is expected since we have considered a flat prior on $\Lambda^2G_1$. Some of these solutions are however not physical: they give a too high value for $M_0$ or $m_s$ (see Ref.~\cite{Biguet2015}). To remain in the physically allowed domain, we shall add an additional prior selecting models for which $M_0 < 400$~MeV (blue contours). We note that our physical solutions contain the one considered in Ref.~\cite{Chanfray_Cchi,chanfray-universe,Chanfray-Schuck} (red star).

\begin{figure}[t]
\centering
\includegraphics[width=0.9\textwidth]{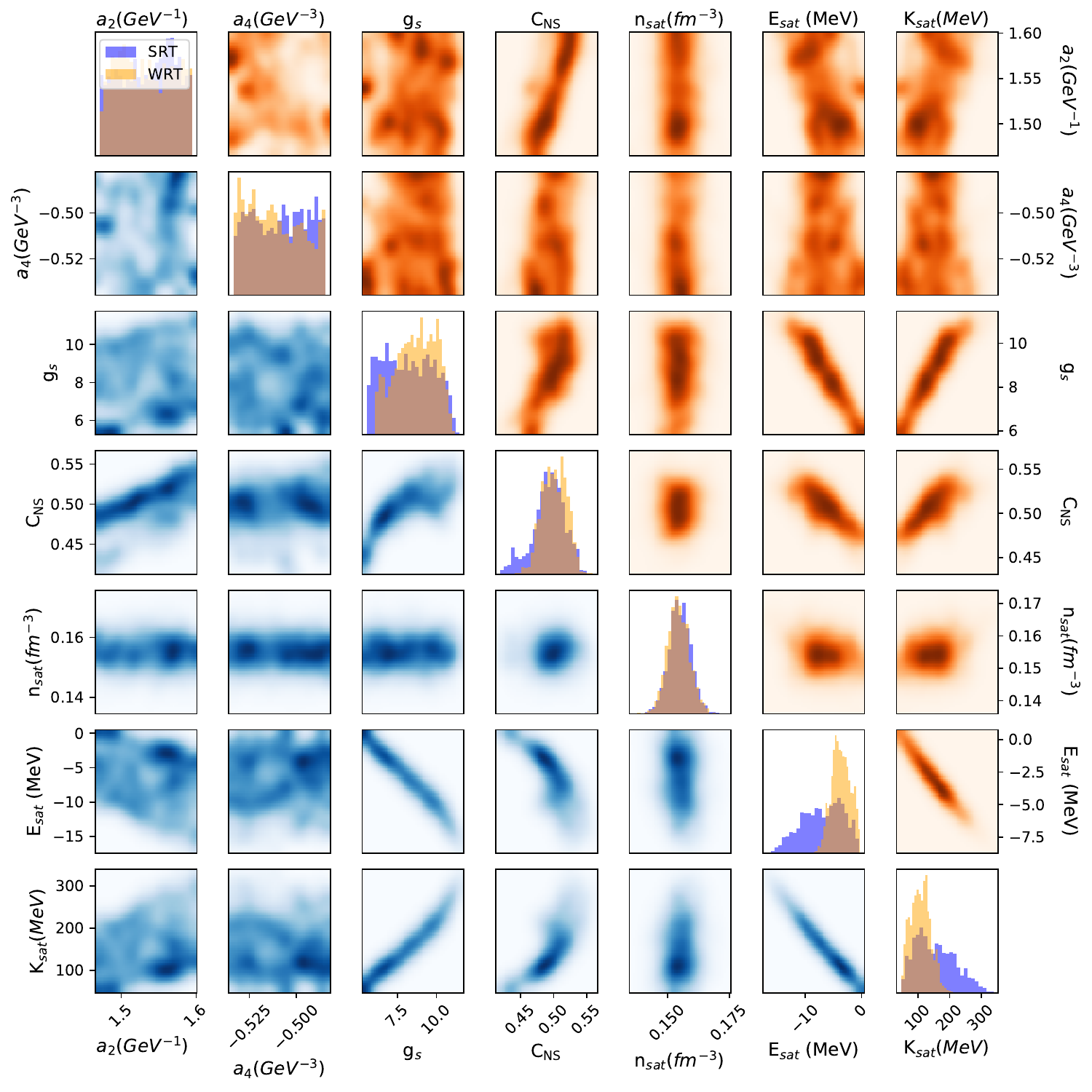}
\caption{Representation of the correlations between the various data and parameters and their individual PDFs for $G_2=0$ in the SRT (blue) and WRT (orange) scenarios.}
\label{fig:corner_noG2}
\end{figure}

In the next section we investigate the full RHF-CC model with additional parameters as well as additional constraints in order to understand which ones of them are able to reproduce the correct saturation properties.

\subsection{The RHF-CC Lagrangian and the nuclear matter properties}

\begin{table*}[t]
\tabcolsep=0.5cm
\def\arraystretch{1.5}
\caption{\label{tab:parameters}%
Model parameters (masses and coupling constants) which are fixed to be constant in the present analysis.}
\begin{tabular}{ccccccc}
\hline
$M_N$ & $m_{\rho}$ & $m_{\delta}$ & $m_{\omega}$ & $g_{\rho}$ & $g_{\delta}$ & $g_A$  \\
MeV & MeV & MeV  & MeV  \\
\hline
938.9 & 779.0 & 984.7 & 783.0 & quark model: $g_{\omega} / 3$ & 1 & 1.25\\
\hline
\end{tabular}
\end{table*}

\begin{table}[t]
\tabcolsep=0.66cm
\def\arraystretch{1.5}
\caption{\label{tab:rhotensor}%
We consider two cases for $\kappa_{\rho}$, the coupling constant of the $\rho_T$: weak $\rho_T$ with $\kappa_{\rho}$=3.7 suggested by the Vector Dominance Model (VDM) \cite{Bhaduri1988}, and strong $\rho_T$ with $\kappa_{\rho}$=6.6 suggested by scattering data~\cite{HOHLER1975210}. }
\begin{tabular}{cccc}
\hline
$\rho_T$ model & & Ref. & $\kappa_{\rho}$ \\
\hline
weak $\rho_T$ & WRT  & \cite{Bhaduri1988} & 3.7 \\
strong $\rho_T$ & SRT &\cite{HOHLER1975210} & 6.6 \\
\hline
\end{tabular}
\end{table}

We now consider the full RHF-CC model considering the conditions imposed by fundamental QCD properties and empirical properties of dense nuclear matter. 

Some parameters in the RHF-CC model are fixed, e.g., the meson masses, the nucleon mass in the vacuum and some coupling constant, see Tab.~\ref{tab:parameters}. Since Eq.~\eqref{eq:msigma} fixes the value for $m_s$, we can determine the value for $g_s$ from the relation~\eqref{eq:a2}. We consider a weak coupling between nucleons and $\delta$ meson by fixing $g_{\delta}=1$. We are only left with $g_\omega$ and $g_\rho$ which are linked together under the simple quark model relation: $g_{\rho} = g_{\omega}/3$. There is therefore only one parameter to fix, e.g., $g_\omega$, which is obtained from the condition to reproduce the nuclear saturation density $n_\sat$. 

In Tab.~\ref{tab:rhotensor}, we give the value considered for the nucleon $\rho$ tensor coupling $f_\rho = g_\rho \kappa_\rho $, according to either the vector dominance model (VDM)~\cite{Bhaduri1988} which we call weak rho tensor (WRT) or to the scattering data~\cite{HOHLER1975210} which we call strong rho tensor (SRT). In this first study, we focus on the SRT and WRT scenarios, and we perform a Bayesian analysis constrained by $n_{\sat}$ as mentioned earlier. The results can be seen in Fig.~\ref{fig:corner_noG2}. 

Fig.~\ref{fig:corner_noG2} represents two corner plots where the lower triangle shows the correlations for the SRT case and the upper triangle the correlations for the WRT case. The marginalised posterior probabilities for these two cases are represented together on the diagonal insets. Note that among all the models which pass the aforementioned constraints, a large number of them do not correctly reproduce the empirical value of the energy $E_\sat$. 

The main differences between the two cases are indeed observed for the posterior distribution function of $E_\sat$ and $K_\sat$. It is clear from Fig.~\ref{fig:corner_noG2} that there is a complication in reproducing $E_\sat$ which is more marked in the WRT case than the SRT case. This can be traced back to the $a_2$ prior: a large value of the binding energy favors large values of $g_s$, the main attractive channel, which also corresponds to a large value of $a_2$ from Eq.~\eqref{eq:a2}. In the WRT scenario, the tensor $\rho$ channel is less attractive and cannot be completed by more attraction in the $s$ channel, which is limited by the prior on $a_2$.

\begin{figure}[t]
\centering
\includegraphics[width=0.9\textwidth]{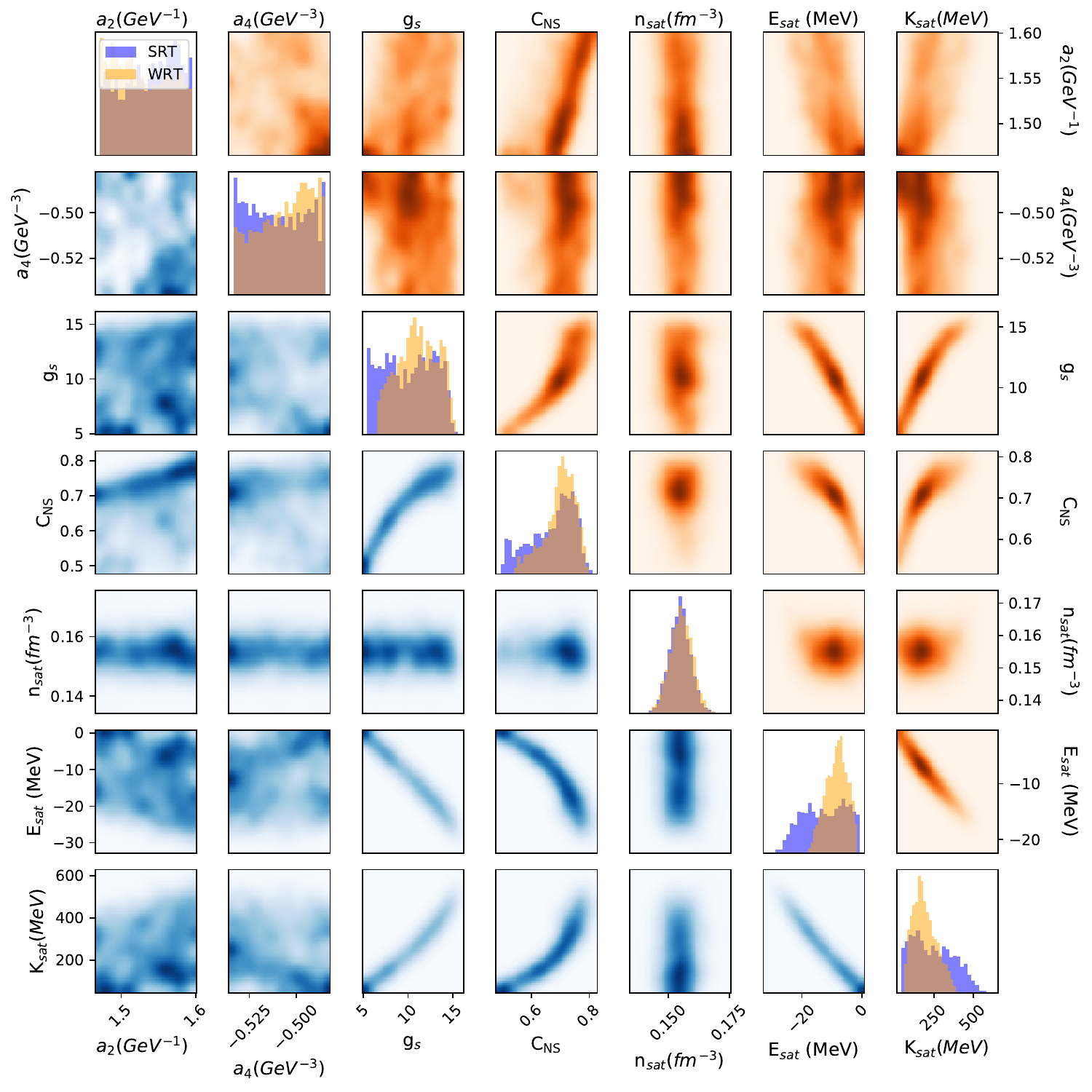}
\caption{Representation of the correlations between the various data and parameters and their individual PDFs for $G_2 = G_1$ in the SRT (blue) and WRT (orange) scenarios.}
\label{fig:corner_G2}
\end{figure}

We perform the same analysis in the case $G_2 = G_1$. Our results are shown in Fig.~\ref{fig:corner_G2}, where the same qualitative features as in Fig.~\ref{fig:corner_noG2} are obtained. Note however a few differences since there are some models which reproduce the empirical values of $E_\sat$ and $K_\sat$, for SRT and WRT. In the WRT case, these models are however a bit marginal.

For the majority of models the saturation energy is however predicted to be different from the empirical expectation. This might be due to the absence of an attractive contribution to the total energy, such as the correlation energy (see Refs.~\cite{Chanfray2007,Toki2009}). So in the present work, we perform a Bayesian inference of the so-called "missing" energy which is required by each model to get the empirical value for the empirical parameters. To do so, we introduce a phenomenological term for the "missing" energy: a term linear in density is constrained by the condition $E_\sat=-15.8\pm0.3$~MeV and a term containing quadratic density dependence is additionally constrained to reproduce the empirical incompressibility modulus $K_\sat=230 \pm20$~MeV~\cite{Margueron2018}. Note that we do not calculate explicitly the correlation energy in the present work.

Furthermore, the parameter $C_\mathrm{NS}$ represented in Figs.~\ref{fig:corner_noG2} and \ref{fig:corner_G2} acquires values which are lower than 1, as expected from the discussion presented in Sec.~\ref{sec:intro}.

\subsection{The "missing" energy}
\label{sec:missing energy}

The "missing" energy is a phenomenological quantity introduced in the present study, function of the density, and added to the energy per particle to reproduce the empirical parameters. It can be calculated microscopically from the correlation energy, for instance, but in this study we develop an approach where it is inferred from the Bayesian analysis. We introduce two analytical forms for the "missing" energy: the first one assumes that it goes linearly with the density, 
\begin{equation}
e^1_\mathrm{m.e.}(n) = \alpha_1 n \, ,
\label{eq:linear}
\end{equation}
and the second one assumes an additional quadratic dependence,
\begin{equation}
e^2_\mathrm{m.e.}(n) = \alpha_2 n + \beta_2 n^2 \, .
\label{eq:quadratic}
\end{equation}
We therefore introduce one (two) additional parameter(s) in the Bayesian analysis, which is (are) controlled by one (two) additional constraints: The condition to reproduce the expected energy per particle at saturation $E_\sat$ (and the incompressibility modulus $K_\sat$).

This study allows us to infer a band of possible "missing" energies, which reproduce the NEPs. In other words, a microscopic calculation of the correlation energy, that we plan to perform in a future study, should fall in our inference bands. Note however that the band inferred by our approach should be larger than the one which would be obtained from a microscopic calculation. 

\begin{figure}[t]
\centering
\includegraphics[width=0.9\textwidth]{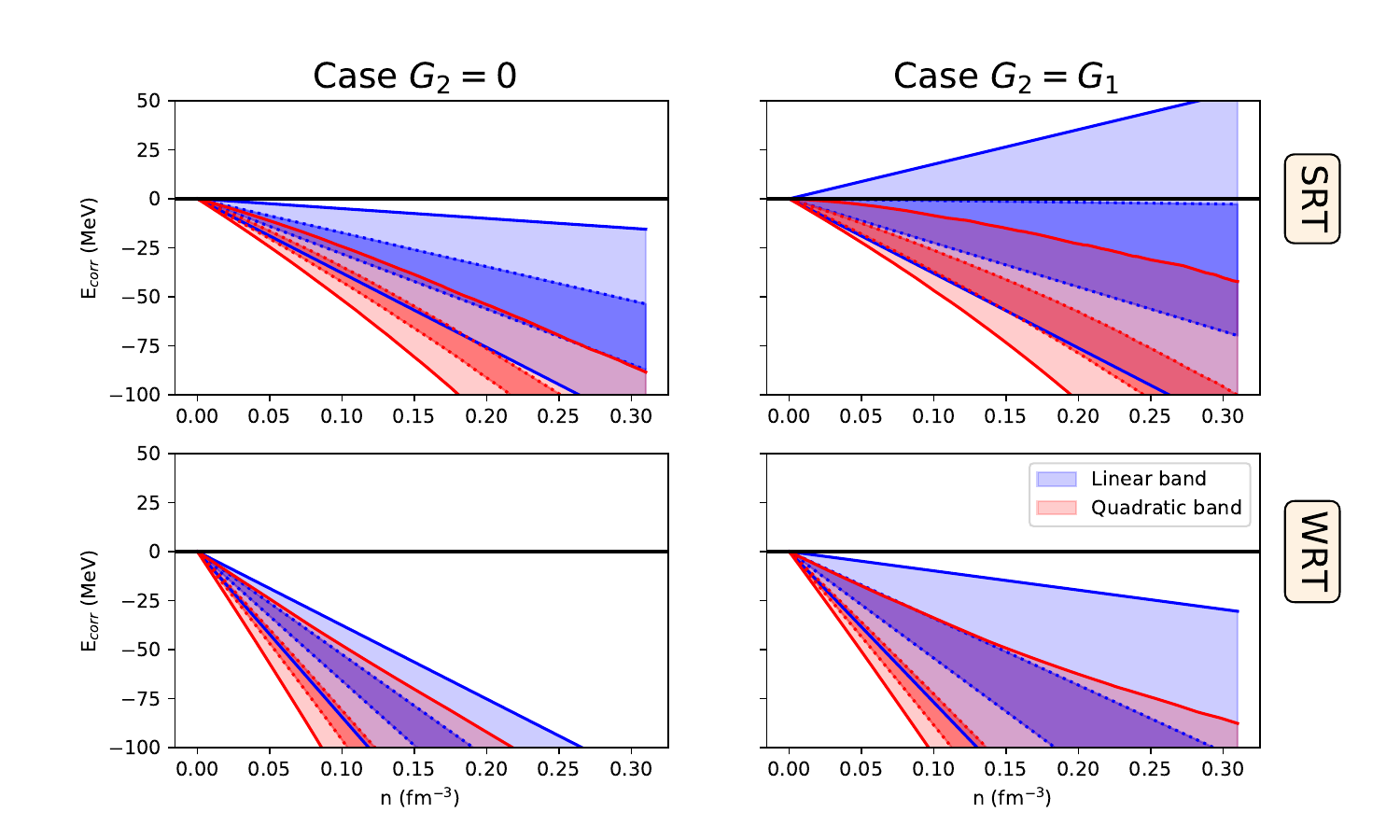}
\caption{Inference on the linear (blue) and quadratic (red) "missing" energy for the SRT (top row) and WRT (bottom row) scenarios.  We show the cases for $G_2 = 0$ on the first column and $G_2=G_1$ on the second column. The bands show the 95\% confidence contour between the straight lines and the 67\% one between the dashed lines.}
\label{fig:inference}
\end{figure}

\begin{figure}[t]
\centering
\includegraphics[width=0.9\textwidth]{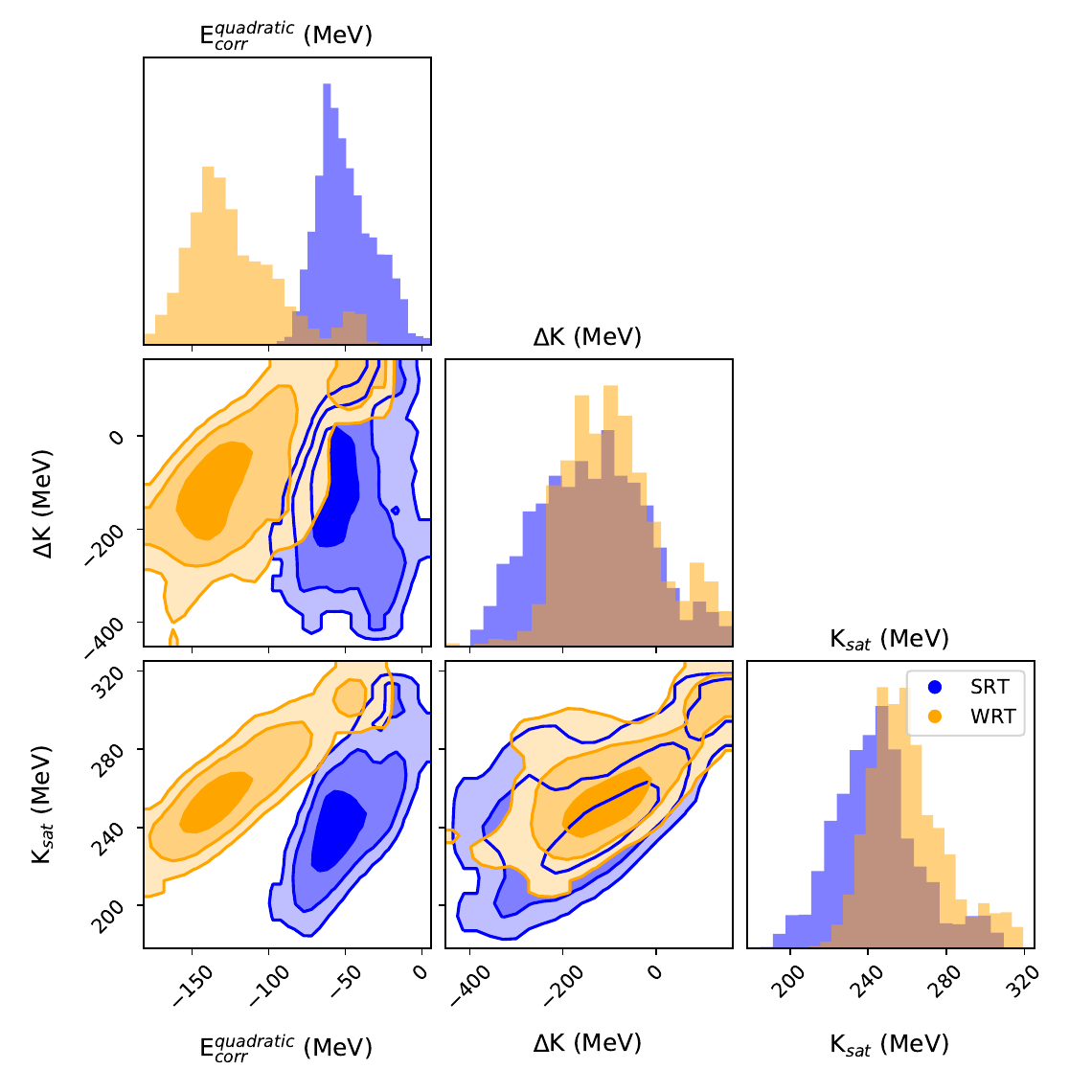}
\caption{Corner plot showing the PDFs and correlations between the value of the quadratic "missing" energy at saturation, the contribution of this energy to the incompressibility $\Delta K$, and $K_{\sat}$ for $G_2=G_1$.}
\label{fig:Ecorrquad_correlation}
\end{figure}

The results of our inference are shown in Fig.~\ref{fig:inference}, where the 67\% and 95\% confidence bands for the cases $G_2=0$ (on the left panels) and $G_2=G_1$ (on the right panels) are shown. We also study the two scenarios: SRT (top panels) and WRT (bottom panels). We first focus on the SRT: for $G_2=0$, we observe that all inferences from this energy, be it a linear or a quadratic density dependence, give negative values, which means that the "missing" energy is predicted to be attractive, which is in agreement with what already was observed in Fig.~\ref{fig:corner_noG2}: the models are located above the expectation value for the saturation energy. The Bayesian inference therefore chose negative values for this "missing" energy. The blue band shows our inference in the case of a linear density dependence for the "missing" energy, see Eq.~\eqref{eq:linear}, while the red band shows the quadratic version, Eq.~\eqref{eq:quadratic}, which is also constrained by the requirement to reproduce $K_{\sat}$. We observe that the constraint on $K_{\sat}$ implies a negative curvature of the "missing" energy (see the peak of the distribution for the "missing" energy's curvature $\Delta K$ represented in Fig.\ref{fig:Ecorrquad_correlation}).

On the second column of Fig.~\ref{fig:corner_noG2}, we study the case with $G_2=G_1$ and SRT for which we recall that some models are able to reproduce the expectation values for the saturation energy $E_\sat$. As a consequence, we expect from our inference that some models predict no contributions from the "missing" energy, or even positive (repulsive) contributions. It is interesting however that in order to also reproduce the correct incompressibility $K_\sat$, only attractive contributions are allowed, be it linear or quadratic. 

For $G_2=G_1$ and the WRT scenarios, all models need negative "missing" energy to reproduce NEPs. The magnitude of the "missing" energy in this case is however globally larger than in the SRT case.  It may even be a bit larger than the Fock contribution to the energy which is of the order of $\sim 20 $ MeV, see Ref.~\cite{Chamseddine} for instance. This may pose question if the "missing" energy breaks the expected hierarchy of many-body contributions, with terms at a given order which shall be smaller in absolute value than their previous order. It would therefore be interesting to check the microscopic calculation of the correlation energy, in particular in the WRT case.

\begin{figure}[t]
\centering
\includegraphics[width=0.9\textwidth]{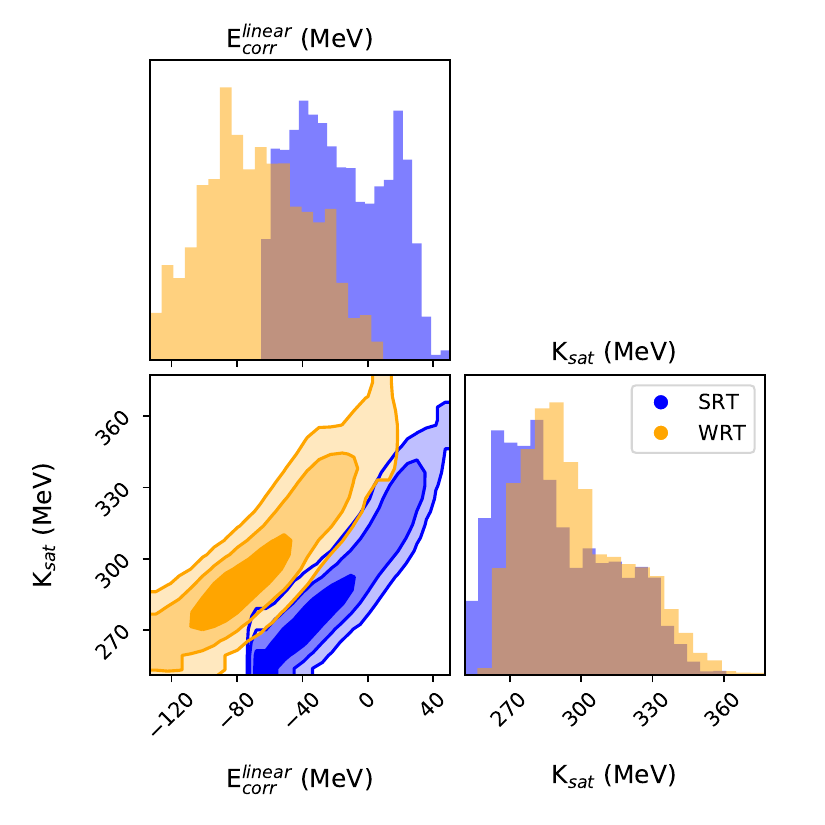}
\caption{Corner plot comparing the PDFs and correlations between the value of the linear "missing" energy at saturation and the incompressibility $K_{\sat}$ for the SRT and WRT cases with $G_2 = G_1$. }
\label{fig:Ecorrlin_correlation}
\end{figure}

To better understand the SRT case, we can look at Figs.~\ref{fig:Ecorrlin_correlation} and \ref{fig:Ecorrquad_correlation}, which show correlations for the case $G_2=G_1$. In Fig.~\ref{fig:Ecorrlin_correlation}, we show the correlation between the value of the linear "missing" energy at saturation $E_{corr}^{linear} = e^1_{m.e}(n_\sat)$ and the incompressibility. There is a positive correlation showing that the larger the "missing" energy, the larger the incompressibility modulus $K_\sat$ with values above the NEP. So in order to match better with the expected value for $K_\sat$, the "missing" energy shall be large and negative (models having $K_\sat \sim 250$ MeV requires $\sim -120$ MeV of "missing" energy).

In Fig.~\ref{fig:Ecorrquad_correlation}, we show that the quadratic "missing" energy has two contributions: one for the saturation energy, corresponding to the value of the "missing" energy at saturation $E_{corr}^{quadratic} = e^2_{m.e}(n_\sat)$, and the other one, the contribution to the incompressibility at saturation which comes from the curvature, i.e $\Delta K = (3n)^2 \frac{\partial^2 e^2_{m.e}(n)}{\partial n^2} \mid_{n=n_\sat}$. As stated in the case before, the energy has to be attractive otherwise it would increase $K_{\sat}$, but interestingly we see that a large majority of models have a negative curvature in order to bring down its value to the experimental one.

The difference between the linear and quadratic contribution is then that for the linear energy to be able to reproduce both $E_{\sat}$ and $K_{\sat}$, a large and negative value of the "missing" energy is required, whereas for the quadratic energy, the value of $K_{\sat}$ is brought down in an interplay between the curvature and the value of the "missing" energy at saturation, which explains why we need smaller attraction.

To summarise, our Bayesian inference for the "missing" energy shows that it is statistically expected to be negative (attractive) with a small negative curvature in the region between $n_\sat$ and $2n_\sat$. Our result provides a motivation for a future study, where we wish to confront our inference to the microscopic calculation of the correlation energy.

\section{Model break-down at high density}
\label{sec:breakdown}

We now explore higher densities and push our model to its limits.
We remark that there is a break down density above which the equations
\begin{eqnarray}
&& V^\prime(\bar s)=-g^*_s n_s \, , \label{eq:sbar} \\ 
&& M^*_D(\mathrm{p}) = M_N + \Sigma_S(\mathrm{p})\, ,  \label{eq:M_D} \\ 
&& \mathbf{p}^* = \mathbf{p} + \mathbf{\tilde{p}}\Sigma_V(\mathrm{p})\, , 
\label{eq: p_star}
\end{eqnarray}
do not have a solution anymore. 
In Eq.~\eqref{eq:sbar}, the effective scalar coupling $g^*_s$ is defined as $g^*_s=\partial M_N(\bar s) / \partial \bar s$ and $n_s$ is the total scalar density defined as
\begin{equation}
n_s = \int\frac{2\,d{\bf p}}{(2\pi)^3}\,\frac{M_D^*(\mathrm{p})}{E^*(\mathrm{p})}\,f(\mathrm{p})\, ,
\end{equation}
with $E^*(p) = \sqrt{M_D^{*2}(\mathrm{p})+\mathbf{p}^{*2}}$ and $f(\mathrm{p}) = \theta(\mathrm{p}_{F} - \mathrm{p})$ is the occupation number for the nucleon characterized by the Fermi momentum $p_{F}$.
In Eqs.~\eqref{eq:M_D}-\eqref{eq: p_star} we introduce the unit momentum vector $\tilde{\bf p} = {\bf p}/\mathrm{p}$, whereas $\Sigma_S(\mathrm{p})$, $\Sigma_0(\mathrm{p})$, and $\Sigma_V(\mathrm{p})$ are the  scalar, time-like and vector-like self-energies, described in Ref.~\cite{Chamseddine} for instance.

\begin{figure}[t]
\centering
\includegraphics[width=0.9\textwidth]{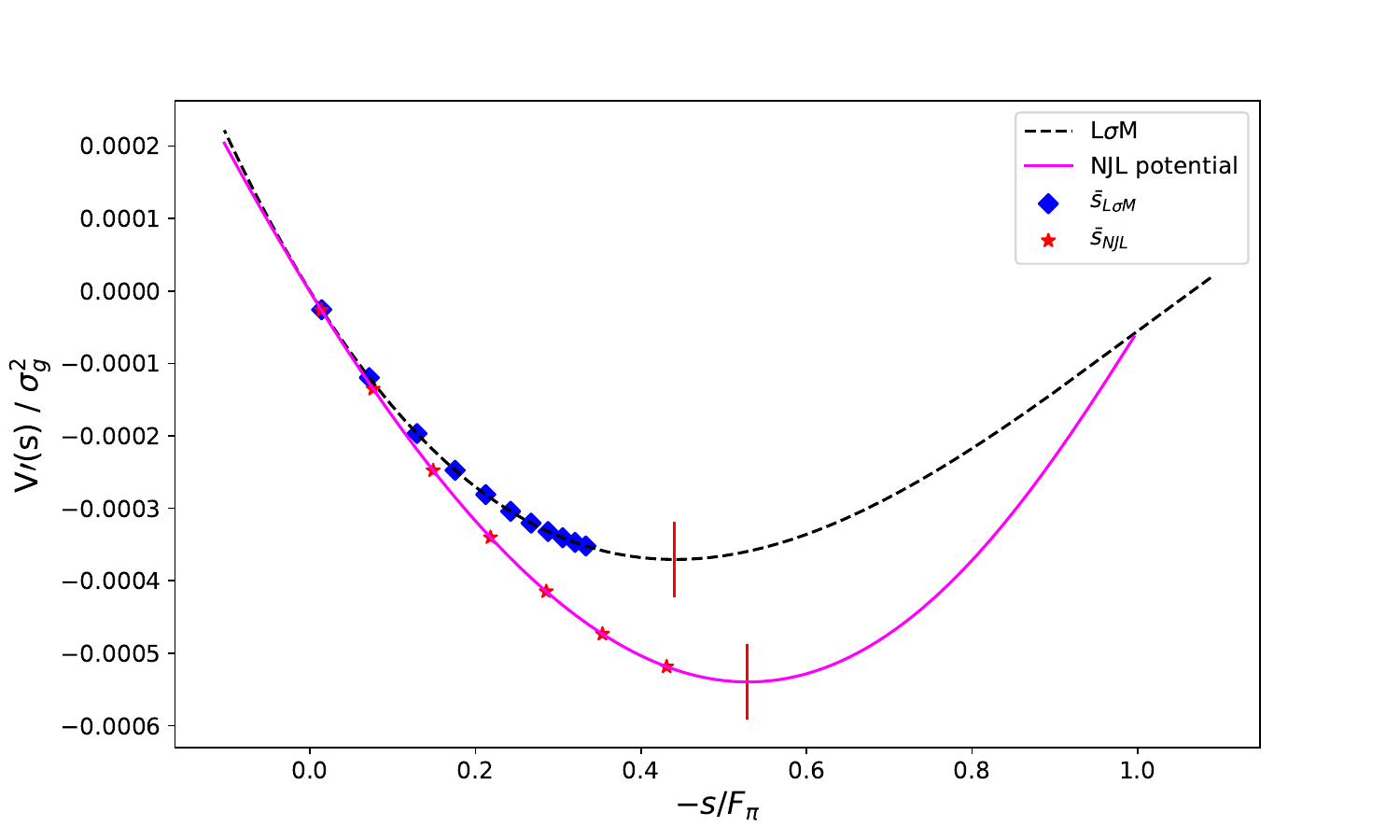}
\caption{Derivative of the potential in units of the string tension as a function of $-\bar{s} / F_\pi$ for the L$\sigma$M (dotted line) and NJL (solid line) models. The points on each curve stand for the solution of Eq.~\eqref{eq:sbar} for the same densities ranging equidistantly from $n = 0.01$~fm$^{-3}$ up to $\sim 0.55$~fm$^{-3}$ for L$\sigma $M and $\sim$ 0.33 fm$^{-3}$ for NJL, and the red bar shows the position of the minimum of $V^\prime(s)$ localising the breakdown density.}
\label{fig:potential_der1}
\end{figure}

The derivative $V^\prime(\bar s)$ is zero for $s=0$ as well as at chiral restoration for $s=-F_\pi$ and it is negative between these two boundaries, see Fig;~\ref{fig:potential_comparison} for instance. There is therefore a value, named ${\bar s}_\mathrm{min}$, for which $V^\prime(\bar s)$ is minimal. It is found to be ${\bar s}_\mathrm{min}=-48.5$ MeV ($-40.5$ MeV) for the same parametrisation of the NJL (L$\sigma$M) potential as in Fig.~\ref{fig:potential_comparison}, see Fig.~\ref{fig:potential_der1}.

We draw in Fig.~\ref{fig:potential_der1} dots corresponding to equidistantly increasing densities for which we search for solutions of the equations of motion~\eqref{eq:sbar}, \eqref{eq:M_D}, and \eqref{eq: p_star}. This is done for the L$\sigma$M and NJL potentials. We observe that for the L$\sigma$M potential, the step in $s$ between two successive densities decrease as one gets closer and closer to the value $s_\mathrm{min}$, while for the NJL potential, this phenomenon is much less visible. As a consequence, the solution of the equation of motion reaches the break-down density at lower density for NJL potential compared to L$\sigma$M potential for the considered parameter set.

\begin{figure}[t]
\centering
\includegraphics[width=0.9\textwidth]{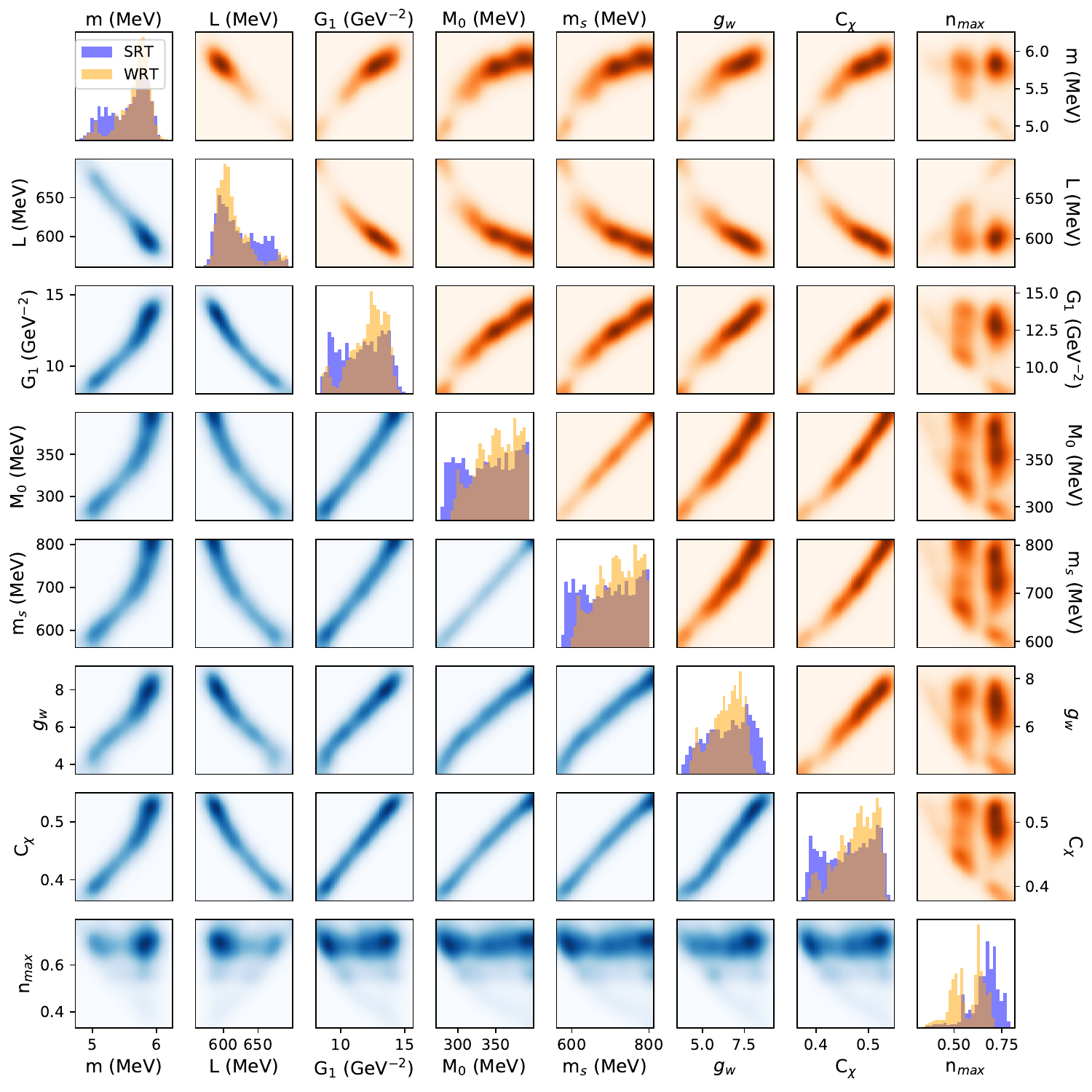}
\caption{Representation of the correlations between the parameters and the breakdown density $n_{max}$ and their individual PDFs for $G_2=0$ in the SRT (blue) and WRT (orange) scenarios.}
\label{fig:nmax_noG2}
\end{figure}

\begin{figure}[t]
\centering
\includegraphics[width=0.9\textwidth]{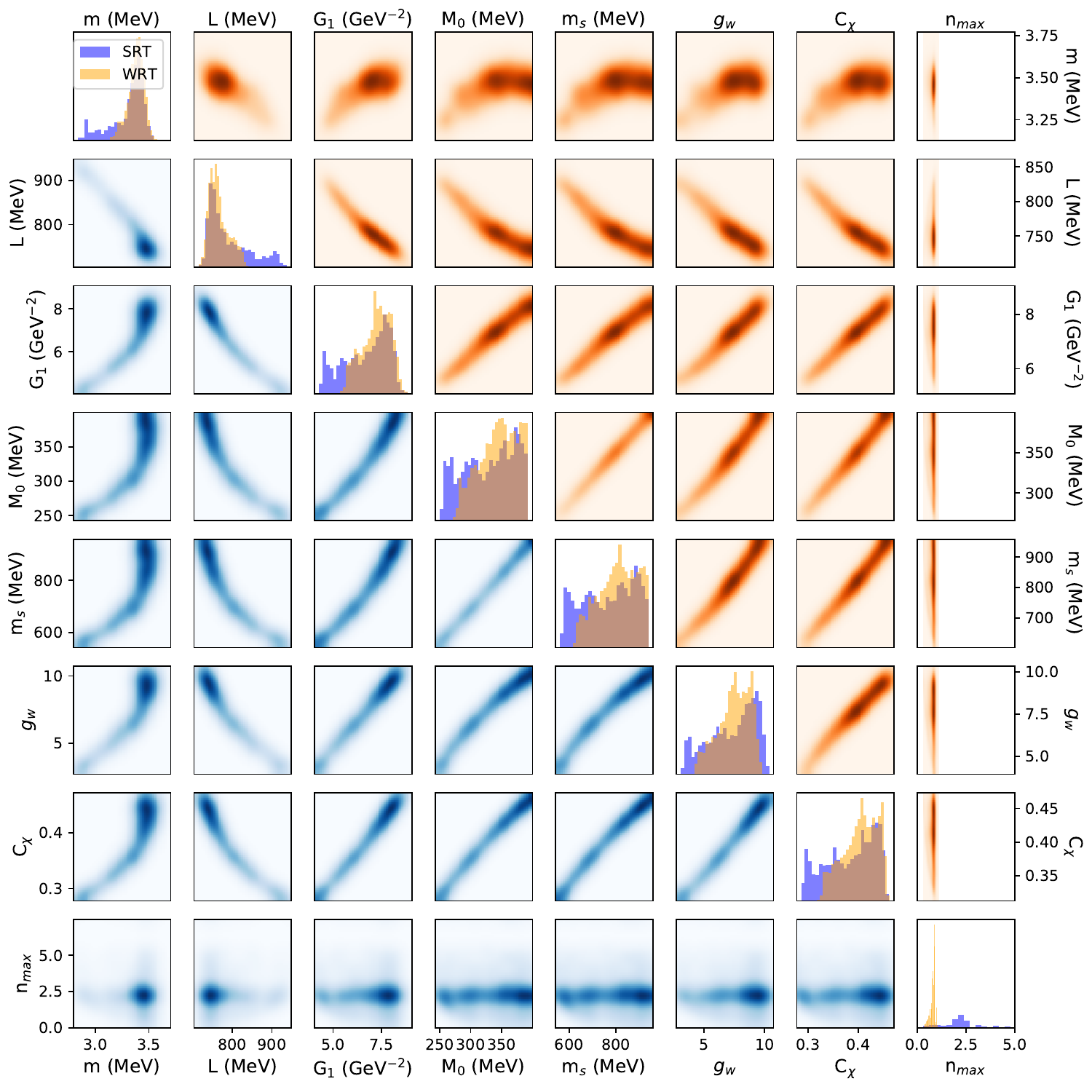}
\caption{Representation of the correlations between the parameters and the breakdown density $n_{max}$ and their individual PDFs for $G_2=G_1$ in the SRT (blue) and WRT (orange) scenarios.}
\label{fig:nmax_G2}
\end{figure}

The value of the break-down density $n_\mathrm{max}$ for which the scalar field is ${\bar s}={\bar s}_\mathrm{min}$ depends however on the model itself. We therefore explore it systematically in a Bayesian way and for the NJL potential, considering the same set of models shown in Figs.~\ref{fig:corner_noG2} and \ref{fig:corner_G2}. Results are shown in Figs.~\ref{fig:nmax_noG2} and \ref{fig:nmax_G2} for the case $G_2=0$ and $G_2=G_1$ respectively. We didn't find any clear correlation between the value of $n_\mathrm{max}$ and the 'model' parameter shown in Figs.~\ref{fig:nmax_noG2} and \ref{fig:nmax_G2}, however we remark that the choice for $\kappa_\rho$ and especially $G_2$ have an important effect on its distribution.

The break down density is of mathematical origin, since it is due to the vanishing of the solution for the equations of motion. It represents therefore the upper density above which our model cannot be applied, and it ranges between $\sim 4$ to $14$ times $n_\sat$. In applying our model to neutron stars, the break down density represent the maximum density of a given EoS. 

\section{Conclusions}

In this paper, we have examined a new chiral potential derived from the NJL model, and its effect on the predictions of the RHF-CC model developed in Ref.~\cite{Chamseddine}. We consider here also the LQCD data as well as the empirical properties to fit our parameters, and we have investigated two sets of NJL potentials, one with the absence of vector interaction ($G_2=0$) and the other one compatible with the $a_1 - \pi$ mixing ($G_2=G_1$).

We have shown that most of the models fail to simultaneously reproduce $E_\sat$ with the required data, given the experimental and LQCD constraints due to a lack of attractive binding energy. This directly lead us to the question of the role of additional contributions to the energy per particle, such as for instance the correlation energy. A phenomenological "missing" energy is added to reproduce the saturation energy with a linear density dependence, or to reproduce the incompressibility modulus with an additional quadratic dependence in the density. We have managed to bring in the extra attraction needed to reproduce the binding energy, and we showed that negative curvatures were additionally required if one wishes to also reproduce the incompressibility. However, for cases where attractive interactions are suppressed, either by the absence of vector interaction which in it turn lowers $m_s$ and thus $g_s$, or by lower values of the attractive tensor interaction, a large contribution is needed to satisfy $E_\sat$ and $K_\sat$.
A more complete model with a microscopic calculation of the correlation energy is envisioned in our future work. We have also focused on symmetric matter, so the effect of the correlation energy on neutron matter is still not known and will also be explored with this microscopic calculation.

Additionally, while pushing the model to high densities we discovered that there is a breakdown density above which the set of equations of motion \eqref{eq:sbar}, \eqref{eq:M_D}, and \eqref{eq: p_star} have no solution. This breakdown is of mathematical origin and is related to the properties of the chiral potential. To our understanding, it is however not related to a physical quantity. No correlation have been found between the model parameters and this breakdown density, which occurs above $\sim 4n_\sat$.

Finally, despite the limitation at high densities, we generated an approach based on RHF-CC with NJL potential, which is anchored in QCD properties and which is able to describe the properties of nuclear matter. With this approach we managed to resolve the conflict between confining models such as the QMC, and LQCD data, all whilst respecting chiral symmetry.

\appendix

\bibliographystyle{spphys}  
\bibliography{biblio}

\end{document}